%% file: main.tex
\documentclass[12pt,a4paper]{article}

\usepackage{lineno}  

\usepackage{graphicx}  

\usepackage{xspace}
\usepackage{color}
\definecolor{mygray}{rgb}{0.5,0.5,0.5}
\usepackage{colortbl}

\usepackage{amsmath}

\usepackage{ifthen} 
\usepackage{rotating}
\usepackage{graphpap}
\usepackage{rotating}
\usepackage{cancel}

\usepackage{multirow}
\usepackage{hyperref}

\usepackage{draftcopy}

\usepackage{verbatim}

\usepackage{longtable} 
\newboolean{pdflatex}
\setboolean{pdflatex}{true} 
\newboolean{uprightparticles}
\setboolean{uprightparticles}{true} 
\usepackage{amssymb}
\usepackage{amsfonts}
\usepackage{upgreek}
\usepackage{cite}

\input{lhcb-symbols-def}

\begin{document}

\renewcommand{\thefootnote}{\fnsymbol{footnote}}
\setcounter{footnote}{1}
\input{title-LHCb-PAPER}

\renewcommand{\thefootnote}{\arabic{footnote}}
\setcounter{footnote}{0}


\pagestyle{plain} 
\setcounter{page}{1}
\pagenumbering{arabic}


\input{introduction}
\input{detector}
\input{evsel}

\input{Nratio}

\input{efficiency}

\input{results}

\bibliographystyle{LHCb}
\bibliography{main,LHCb-PAPER,references1}

\end{document}

%% file: lhcb-symbols-def.tex
\def\lhcb {\mbox{LHCb}\xspace}
\def\ux85 {\mbox{UX85}\xspace}

\def\belle  {\mbox{Belle}\xspace}

\def\cdf    {\mbox{CDF}\xspace}
\def\dzero  {\mbox{D0}\xspace}

\def\rich   {RICH\xspace}

\ifthenelse{\boolean{uprightparticles}}%
{
 
 \def\Pgamma      {\ensuremath{\upgamma}\xspace}

 \def\Peta        {\ensuremath{\upeta}\xspace}

 \def\Pmu         {\ensuremath{\upmu}\xspace}

 \def\Ppi         {\ensuremath{\uppi}\xspace}                 
                  
 \def\Prho        {\ensuremath{\uprho}\xspace}

 \def\Ppsi        {\ensuremath{\uppsi}\xspace}

 \def\PDelta      {\ensuremath{\Delta}\xspace}                 
 \def\PXi         {\ensuremath{\Xi}\xspace}                 
 \def\PLambda     {\ensuremath{\Lambda}\xspace}                 
 \def\PSigma      {\ensuremath{\Sigma}\xspace}                 
 \def\POmega      {\ensuremath{\Omega}\xspace}                 
 \def\PUpsilon    {\ensuremath{\Upsilon}\xspace}

 \def\PB      {\ensuremath{\mathrm{B}}\xspace}                 
                  
 \def\PD      {\ensuremath{\mathrm{D}}\xspace}

 \def\PJ      {\ensuremath{\mathrm{J}}\xspace}                 
 \def\PK      {\ensuremath{\mathrm{K}}\xspace}

 \def\Pb      {\ensuremath{\mathrm{b}}\xspace}                 
 \def\Pc      {\ensuremath{\mathrm{c}}\xspace}                 
                  
 \def\Pe      {\ensuremath{\mathrm{e}}\xspace}

 \def\Pi      {\ensuremath{\mathrm{i}}\xspace}

 \def\Pp      {\ensuremath{\mathrm{p}}\xspace}

 \def\Ps      {\ensuremath{\mathrm{s}}\xspace}

}
{
 
 \def\Pgamma      {\ensuremath{\gamma}\xspace}

 \def\Peta        {\ensuremath{\eta}\xspace}

 \def\Pmu         {\ensuremath{\mu}\xspace}

 \def\Ppi         {\ensuremath{\pi}\xspace}                 
                  
 \def\Prho        {\ensuremath{\rho}\xspace}

 \def\Ppsi        {\ensuremath{\psi}\xspace}                 
                  
 \mathchardef\PDelta="7101
 \mathchardef\PXi="7104
 \mathchardef\PLambda="7103
 \mathchardef\PSigma="7106
 \mathchardef\POmega="710A
 \mathchardef\PUpsilon="7107
                  
 \def\PB      {\ensuremath{B}\xspace}                 
                  
 \def\PD      {\ensuremath{D}\xspace}

 \def\PJ      {\ensuremath{J}\xspace}                 
 \def\PK      {\ensuremath{K}\xspace}

 \def\Pb      {\ensuremath{b}\xspace}                 
 \def\Pc      {\ensuremath{c}\xspace}                 
                  
 \def\Pe      {\ensuremath{e}\xspace}

 \def\Pi      {\ensuremath{i}\xspace}

 \def\Pp      {\ensuremath{p}\xspace}

 \def\Ps      {\ensuremath{s}\xspace}

}

\def\en         {\ensuremath{\Pe^-}\xspace}   
\def\ep         {\ensuremath{\Pe^+}\xspace}

\def\mup        {\ensuremath{\Pmu^+}\xspace}
\def\muM        {\ensuremath{\Pmu^-}\xspace} 
\def\mumu       {\ensuremath{\Pmu^+\Pmu^-}\xspace}

\def\squark{\ensuremath{\Ps}\xspace}

\def\c     {\ensuremath{\Pc}\xspace}

\def\b     {\ensuremath{\Pb}\xspace}

\def\pion  {\ensuremath{\Ppi}\xspace}
\def\piz   {\ensuremath{\pion^0}\xspace}

\def\pip   {\ensuremath{\pion^+}\xspace}
\def\pim   {\ensuremath{\pion^-}\xspace}
\def\pipi  {\ensuremath{\pion^+\pion^-}\xspace}

\def\rhomeson  {\ensuremath{\Prho^{0}(770)}\xspace}

\def\kaon  {\ensuremath{\PK}\xspace}
  \def\Kbar  {\kern 0.2em\overline{\kern -0.2em \PK}{}\xspace}

\def\Kz    {\ensuremath{\kaon^0}\xspace}
\def\Kzb   {\ensuremath{\Kbar^0}\xspace}
\def\KzKzb {\ensuremath{\Kz \kern -0.16em \Kzb}\xspace}
\def\Kp    {\ensuremath{\kaon^+}\xspace}
\def\Km    {\ensuremath{\kaon^-}\xspace}

\def\KpKm  {\ensuremath{\Kp \kern -0.16em \Km}\xspace}
\def\KS    {\ensuremath{\kaon^0_{\rm\scriptscriptstyle S}}\xspace} 
 
\def\Kstarz  {\ensuremath{\kaon^{*0}}\xspace}

\newcommand{\etapr}{\ensuremath{\Peta^{\prime}}\xspace}

  \def\Dbar    {\kern 0.2em\overline{\kern -0.2em \PD}{}\xspace}
\def\D       {\ensuremath{\PD}\xspace}

\def\Dz      {\ensuremath{\D^0}\xspace}
\def\Dzb     {\ensuremath{\Dbar^0}\xspace}
\def\DzDzb   {\ensuremath{\Dz {\kern -0.16em \Dzb}}\xspace}
\def\Dp      {\ensuremath{\D^+}\xspace}
\def\Dm      {\ensuremath{\D^-}\xspace}

\def\DpDm    {\ensuremath{\Dp {\kern -0.16em \Dm}}\xspace}

\def\B       {\ensuremath{\PB}\xspace}
\def\Bbar    {\ensuremath{\kern 0.18em\overline{\kern -0.18em \PB}{}}\xspace}

\def\Bd      {\ensuremath{\B^0}\xspace}
\def\Bs      {\ensuremath{\B^0_\squark}\xspace}
\def\Bsb     {\ensuremath{\Bbar^0_\squark}\xspace}

\def\BorBs   {\ensuremath{\B^{0}_{\mathrm{(s)}}}\xspace}

\def\jpsi     {\ensuremath{{\PJ\mskip -3mu/\mskip -2mu\Ppsi\mskip 2mu}}\xspace}
\def\psitwos  {\ensuremath{\Ppsi{(2\mathrm{S})}}\xspace}
\def\jporptwo {\ensuremath{\jpsi(\psitwos)}\xspace}

  \def\Y#1S{\ensuremath{\PUpsilon{(#1S)}}\xspace}

\def\proton      {\ensuremath{\Pp}}

\def\BF         {{\ensuremath{\cal B}\xspace}}

\def\BR         {\BF}

\def\to                 {\ensuremath{\rightarrow}\xspace}

\def\CP                {\ensuremath{C\!P}\xspace}

\def\AT#1     {\ensuremath{A_{\mathrm{T}}^{#1}}\xspace}           

\def\C#1      {\ensuremath{\mathcal{C}_{#1}}\xspace}                       
\def\Cp#1     {\ensuremath{\mathcal{C}_{#1}^{'}}\xspace}                    
\def\Ceff#1   {\ensuremath{\mathcal{C}_{#1}^{\mathrm{(eff)}}}\xspace}        
\def\Cpeff#1  {\ensuremath{\mathcal{C}_{#1}^{'\mathrm{(eff)}}}\xspace}       
\def\Ope#1    {\ensuremath{\mathcal{O}_{#1}}\xspace}                       
\def\Opep#1   {\ensuremath{\mathcal{O}_{#1}^{'}}\xspace}                    

\newcommand{\tev}{\ensuremath{\mathrm{\,Te\kern -0.1em V}}\xspace}
\newcommand{\gev}{\ensuremath{\mathrm{\,Ge\kern -0.1em V}}\xspace}
\newcommand{\mev}{\ensuremath{\mathrm{\,Me\kern -0.1em V}}\xspace}
\newcommand{\kev}{\ensuremath{\mathrm{\,ke\kern -0.1em V}}\xspace}
\newcommand{\ev}{\ensuremath{\mathrm{\,e\kern -0.1em V}}\xspace}
\newcommand{\gevc}{\ensuremath{{\mathrm{\,Ge\kern -0.1em V\!/}c}}\xspace}
\newcommand{\mevc}{\ensuremath{{\mathrm{\,Me\kern -0.1em V\!/}c}}\xspace}
\newcommand{\gevcc}{\ensuremath{{\mathrm{\,Ge\kern -0.1em V\!/}c^2}}\xspace}
\newcommand{\gevgevcccc}{\ensuremath{{\mathrm{\,Ge\kern -0.1em V^2\!/}c^4}}\xspace}
\newcommand{\mevcc}{\ensuremath{{\mathrm{\,Me\kern -0.1em V\!/}c^2}}\xspace}

\def\mum  {\ensuremath{\,\upmu\rm m}\xspace}

\def\invfb   {\ensuremath{\mbox{\,fb}^{-1}}\xspace}

\newcommand{\stat}{\ensuremath{\mathrm{(stat)}}\xspace}
\newcommand{\syst}{\ensuremath{\mathrm{(syst)}}\xspace}

\newcommand{\chisq}{\ensuremath{\chi^2}\xspace}

\def\gsim{{~\raise.15em\hbox{$>$}\kern-.85em
          \lower.35em\hbox{$\sim$}~}\xspace}
\def\lsim{{~\raise.15em\hbox{$<$}\kern-.85em
          \lower.35em\hbox{$\sim$}~}\xspace}

\def\sPlot{\mbox{\em sPlot}}

\def\sqs   {\ensuremath{\protect\sqrt{s}}\xspace}

\def\pt         {\mbox{$p_{\rm T}$}\xspace}

\def\evtgen     {\mbox{\textsc{EvtGen}}\xspace}
\def\pythia     {\mbox{\textsc{Pythia}}\xspace}

\def\geant      {\mbox{\textsc{Geant4}}\xspace}

\def\photos     {\mbox{\textsc{Photos}}\xspace}

\def\tell1  {TELL1\xspace}
\def\ukl1   {UKL1\xspace}



%% file: title-LHCb-PAPER.tex
\begin{titlepage}
\pagenumbering{roman}

\vspace*{-1.5cm}
\centerline{\large EUROPEAN ORGANIZATION FOR NUCLEAR RESEARCH (CERN)}
\vspace*{1.5cm}
\hspace*{-0.5cm}
\begin{tabular*}{\linewidth}{lc@{\extracolsep{\fill}}r}
\ifthenelse{\boolean{pdflatex}}
{\vspace*{-2.7cm}\mbox{\!\!\!\includegraphics[width=.14\textwidth]{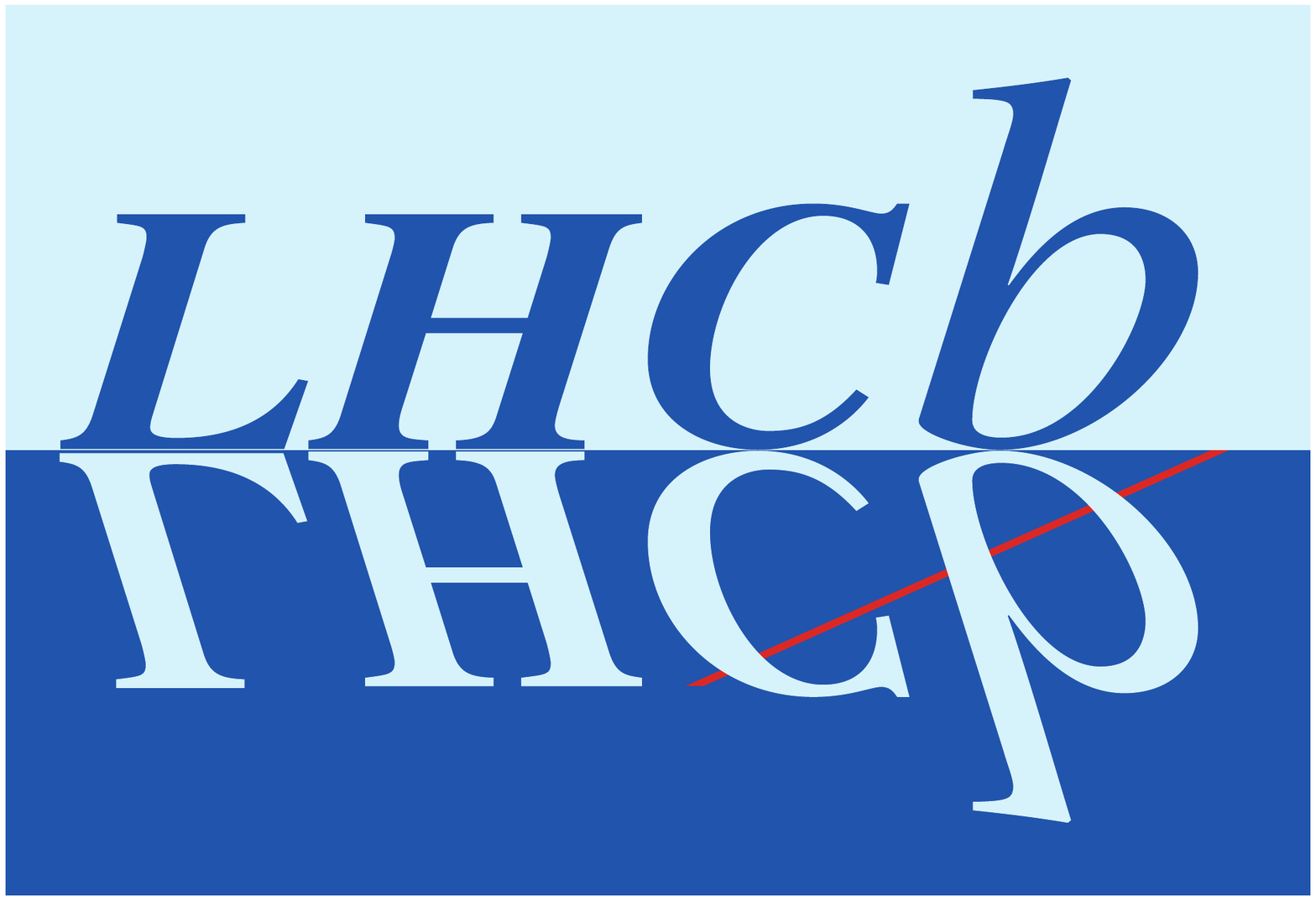}} & &}%
{\vspace*{-1.2cm}\mbox{\!\!\!\includegraphics[width=.12\textwidth]{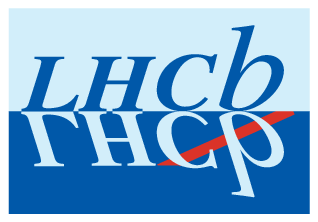}} & &}%
\\
 & & CERN-PH-EP-2013-024 \\  
 & & LHCb-PAPER-2012-053 \\  
 & & 25 February 2013 \\ 
 & & \\
\end{tabular*}

\vspace*{1.2cm}

{\bf\boldmath\huge
\begin{center}
    Observations of 
    \boldmath$\Bs\to\psitwos\Peta$    and 
    \boldmath$\BorBs\to\psitwos\pipi$ decays
\end{center}
}

\vspace*{0.7cm}

\begin{center}
The LHCb collaboration\footnote{Authors are listed on the following pages.}
\end{center}

\vspace{\fill}

\begin{abstract}
  \noindent
First observations of the $\Bs\to\psitwos\Peta$, 
$\Bd\to\psitwos\pipi$ and $\Bs\to\psitwos\pipi$ decays are made 
using a dataset corresponding to an integrated luminosity 
of 1.0\invfb collected by the LHCb experiment in proton-proton 
collisions at a centre-of-mass energy of $\sqs=7\tev$.
The ratios of the branching fractions of each of the \psitwos modes with respect to 
the corresponding \jpsi decays are 
\begin{align*}
\frac{\BR(\Bs\to \psitwos \Peta)    }{\BR(\Bs\to \jpsi \Peta)}    &=0.83\pm0.14\,\stat\pm0.12\,\syst\pm0.02\,(\BR), \\
\frac{\BR(\Bd\to \psitwos \pip \pim)}{\BR(\Bd\to \jpsi \pip \pim)}&=0.56\pm0.07\,\stat\pm0.05\,\syst\pm0.01\,(\BR), \\
\frac{\BR(\Bs\to \psitwos \pip \pim)}{\BR(\Bs\to \jpsi \pip \pim)}&=0.34\pm0.04\,\stat\pm0.03\,\syst\pm0.01\,(\BR), 
\end{align*}
where the third uncertainty corresponds to the uncertainties of the dilepton branching 
fractions of the $\jpsi$ and $\psitwos$ meson decays. 
\end{abstract}

\vspace*{0.8cm}

\begin{center}
  Submitted to Nucl. Phys. B
\end{center}
\vspace{\fill}
{\footnotesize 
\centerline{\copyright~CERN on behalf of the \lhcb collaboration, license \href{http://creativecommons.org/licenses/by/3.0/}{CC-BY-3.0}.}}
\vspace*{2mm}

\end{titlepage}


\newpage
\setcounter{page}{2}
\mbox{~}
\input{LHCb_authorlist.tex}

\cleardoublepage

%% file: LHCb_authorlist.tex
\centerline{\large\bf LHCb collaboration}
\begin{flushleft}
\small
R.~Aaij$^{40}$, 
C.~Abellan~Beteta$^{35,n}$, 
B.~Adeva$^{36}$, 
M.~Adinolfi$^{45}$, 
C.~Adrover$^{6}$, 
A.~Affolder$^{51}$, 
Z.~Ajaltouni$^{5}$, 
J.~Albrecht$^{9}$, 
F.~Alessio$^{37}$, 
M.~Alexander$^{50}$
% ...

{\footnotesize \it
% ...
$ ^{5}$Clermont Universit\'{e}, Universit\'{e} Blaise Pascal, CNRS/IN2P3, LPC, Clermont-Ferrand, France\\
$ ^{6}$CPPM, Aix-Marseille Universit\'{e}, CNRS/IN2P3, Marseille, France\\
% ...
$ ^{9}$Fakult\"{a}t Physik, Technische Universit\"{a}t Dortmund, Dortmund, Germany\\
% ...
$ ^{35}$Universitat de Barcelona, Barcelona, Spain\\
$ ^{36}$Universidad de Santiago de Compostela, Santiago de Compostela, Spain\\
$ ^{37}$European Organization for Nuclear Research (CERN), Geneva, Switzerland\\
% ...
$ ^{40}$Nikhef National Institute for Subatomic Physics, Amsterdam, The Netherlands\\
% ...
$ ^{45}$H.H. Wills Physics Laboratory, University of Bristol, Bristol, United Kingdom\\
% ...
$ ^{50}$School of Physics and Astronomy, University of Glasgow, Glasgow, United Kingdom\\
$ ^{51}$Oliver Lodge Laboratory, University of Liverpool, Liverpool, United Kingdom\\
% ...
$ ^{n}$LIFAELS, La Salle, Universitat Ramon Llull, Barcelona, Spain\\
% ...
}
\end{flushleft}

%% file: introduction.tex
\section{Introduction}
\label{sec:Introduction}

Decays of B~mesons containing a charmonium resonance, \jpsi or \psitwos,
in the final state play a crucial role in the study of $\CP$~violation 
and in the precise measurement of neutral B meson mixing parameters.

The $\Bs\to\jpsi\Peta$ decay was observed by the \belle collaboration and the 
branching fraction was measured to be
$\BR(\Bs\to\jpsi\Peta) = (5.10 \pm 0.50 \pm 0.25\,^{+1.14}_{-0.79})\times 10^{-4}$~\cite{Belle_article}, 
where the first uncertainty is statistical, the second systematic 
and the third due to the uncertainty in the number of produced $\Bs\Bsb$ pairs. 
This decay has also recently been reported by \lhcb, including the decay 
$\Bs\to\jpsi\etapr$~\cite{LHCb-PAPER-2012-022}.

The $\BorBs\to\jpsi\pipi$~decays,
where \BorBs denotes a \Bd or \Bs meson,
have been studied previously and the \pipi final 
states are found to comprise the decay products of 
the $\rhomeson$ and $\mathrm{f_2}(1270)$ mesons in case of 
\Bd~decays and of $\mathrm{f_0}(980)$ and $\mathrm{f_0}(1370)$~mesons 
in case of \Bs~decays~\cite{LHCb-PAPER-2011-002,LHCb-PAPER-2012-045,LHCb-PAPER-2012-005}. 
The \Bs~modes have been used to measure mixing-induced $\CP$ 
violation~\cite{LHCb-PAPER-2011-031,LHCb-PAPER-2012-006}. 
The decays $\Bs\to\psitwos\Peta$ and $\BorBs\to\psitwos\pipi$~have not previously been studied. 

The relative branching fractions of \Bd and 
\Bs~mesons into final 
states containing $\jpsi$ and $\psitwos$ mesons have been studied by several 
experiments (\cdf~\cite{CDF_art,Abulencia:2006jp}, \dzero~\cite{D0_art} and \lhcb~\cite{LHCb-PAPER-2012-010}).
In this paper, measurements of the branching fraction ratios of \BorBs mesons 
decaying to $\psitwos\mathrm{X^{0}}$ and $\jpsi \mathrm{X^{0}}$ are reported, 
where  $\mathrm{X^{0}}$~denotes either an $\Peta$~meson or a $\pipi$~system.
Charge conjugate decays are implicitly included. 
The analysis presented here is based on a data sample corresponding to an 
integrated luminosity of 1.0\invfb collected with the \lhcb 
detector during $2011$ in $\proton\proton$ collisions at a centre-of-mass energy of $\sqs=7\tev$.

%% file: detector.tex
\section{LHCb detector}
\label{sec:Detector}

The \lhcb detector~\cite{Alves:2008zz} is a single-arm forward
spectrometer covering the \mbox{pseudorapidity} range $2<\eta <5$,
designed for the study of particles containing \b or \c 
quarks. The detector includes a high precision tracking system
consisting of a silicon-strip vertex detector surrounding the $\proton\proton$ 
interaction region, a large-area silicon-strip detector located
upstream of a dipole magnet with a bending power of about
$4{\rm\,Tm}$, and three stations of silicon-strip detectors and straw
drift tubes placed downstream. The combined tracking system has 
momentum resolution $\Delta p/p$ that varies from 0.4\% at 5\gevc to
0.6\% at 100\gevc, and impact parameter resolution of 20\mum for
tracks with high transverse momentum~(\pt). Charged hadrons are identified
using two ring-imaging Cherenkov detectors. Photon, electron and
hadron candidates are identified by a calorimeter system consisting of
scintillating-pad and preshower detectors, an electromagnetic
calorimeter and a hadronic calorimeter. Muons are identified by a
system composed of alternating layers of iron and multiwire
proportional chambers. 

The trigger~\cite{Aaij:2012me} consists of a 
hardware stage, based on information from the calorimeter and muon 
systems, followed by a software stage where a full event reconstruction is applied. 
Candidate events are first required to pass a hardware trigger 
which selects muons with a transverse momentum, $\pt>1.48\gevc$. In 
the subsequent software trigger, at least 
one of the final state particles is required to have both 
$\pt>0.8\gevc$ and impact parameter $>100\mum$ with respect to all 
of the primary $\proton\proton$ interaction vertices~(PVs) in the 
event. Finally, two or more of the final state 
particles are required to form a vertex which is significantly 
displaced from the PVs.

For the simulation, $\proton\proton$ collisions are generated using
\pythia~6.4~\cite{Sjostrand:2006za} with a specific \lhcb~configuration~\cite{LHCb-PROC-2010-056}.  
Decays of hadronic particles
are described by \evtgen~\cite{Lange:2001uf} in which final state
radiation is generated using \photos~\cite{Golonka:2005pn}. 
The interaction of the generated particles with the detector and its
response are implemented using the \geant toolkit~\cite{Agostinelli:2002hh,Allison:2006ve} 
as described in Ref.~\cite{LHCb-PROC-2011-006}.

%% file: evsel.tex
\section{Event selection}
\label{sec:EventSelection}

The decays  $\BorBs\to\Ppsi\Peta$ and $\BorBs\to\Ppsi\pipi$, where 
$\Ppsi$ denotes \jpsi~or \psitwos,  are reconstructed using 
$\Ppsi\to\mumu$ and $\Peta\to\Pgamma\Pgamma$~decay modes. 
Pairs of oppositely-charged tracks 
identified as muons, each having $\pt>0.55\gevc$ and originating 
from a common vertex, are combined to form $\Ppsi\to\mumu$ candidates. Track quality 
is ensured by requiring the $\chisq$ per number of degrees of freedom 
($\chi^2/\mathrm{ndf}$) provided by the track fit to be less than~5.
Well identified muons are selected by requiring that
the difference in logarithms of the global likelihood of the muon hypothesis, 
$\Delta\log \mathcal{L}_{\mu\mathrm{h}}$~\cite{Muon:performance}, provided by the 
particle identification detectors,
with respect to the hadron hypothesis is larger than zero. 
The fit of the common two-prong vertex is required to satisfy 
$\chisq/\mathrm{ndf}<20$. The vertex is deemed to be well separated 
from the reconstructed primary vertex of the proton-proton 
interaction by requiring the decay length significance
to be larger than three. Finally, the invariant mass of the dimuon combination 
is required to be between 3.020 and 3.135$\gevcc$ for $\jpsi$ candidates and between 
3.597 and 3.730$\gevcc$ for $\psitwos$ candidates. These correspond to [--5$\sigma$; 
3$\sigma$] intervals around the nominal masses to accomodate 
QED radiation.

The pions are required to have $\pt>0.25\gevc$ and 
an impact parameter $\chi^2$, defined as the difference 
between the $\chi^2$ of the PV formed with and without 
the considered track, larger than 9. When more that one
PV is reconstructed, the smallest value of 
impact parameter $\chi^2$ is chosen. 
In addition,  to suppress contamination from  kaons, 
the difference between the logarithms of likelihoods of 
the pion and kaon hypotheses, $\Delta\log \mathcal{L}_{\Ppi\kaon}$~\cite{arxiv:1211-6759}, 
provided by the \rich detectors, has to be larger than zero. 

Photons are selected from neutral clusters in the electromagnetic 
calorimeter with transverse energy in excess of \mbox{$0.4\gev$.} 
The \mbox{$\Peta\to\Pgamma\Pgamma$}~candidates are reconstructed as
diphoton combinations with an invariant mass within
$\pm70\mevcc$ of the $\Peta$~mass~\cite{PDG2012}.
To suppress the large combinatorial background 
from the decays of neutral pions, photons that form a
$\piz\rightarrow\Pgamma\Pgamma$~candidate with 
invariant mass within $\pm25\mevcc$~of the $\piz$~mass 
are not used to reconstruct $\Peta\to\Pgamma\Pgamma$~candidates.

The $\BorBs$~candidates are formed from $\Ppsi\mathrm{X^{0}}$~combinations.
In the $\Ppsi\Peta$ case an additional requirement~\mbox{$\pt(\Peta)>2.5\gevc$} is 
applied to reduce combinatorial background. 
To improve the invariant mass resolution a kinematic fit~\cite{Hulsbergen:2005pu} 
is performed. In this fit, constraints are applied on the known masses~\cite{PDG2012}
of intermediate resonances, and it is also required that the
candidate's momentum vector points to the associated primary vertex. The 
$\chi^2/\mathrm{ndf}$ for this fit is required to be less than 5. 
Finally, the decay time,  $ct$, of the $\BorBs$~candidate, 
calculated with respect  to the primary vertex, 
is required to be in excess of~$150$\mum.

%% file: Nratio.tex
\newcommand {\lhcbxpos} {57}
\newcommand {\lhcbXpos} {137}
\newcommand {\lhcbypos} {47}

\section[Observation of the           $\Bs\to\psitwos\Peta$ decay]
        {Observation of the  \boldmath$\Bs\to\psitwos\Peta$ decay}
\label{sec:Nratio}

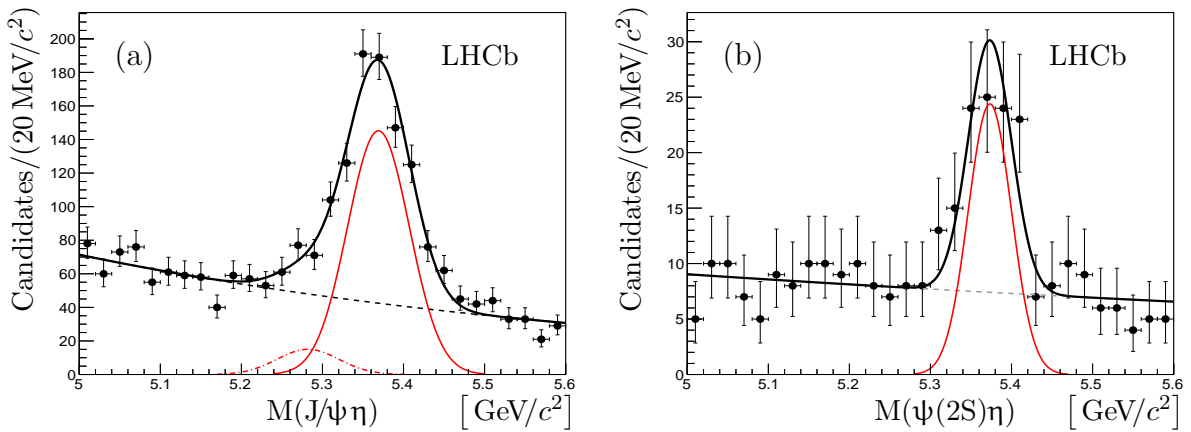
\begin{figure}[ht]
  \setlength{\unitlength}{1mm}
  \centering
  \begin{picture}(160,60)
    \put(0,0){
      \includegraphics*[width=80mm,height=60mm%
      ]{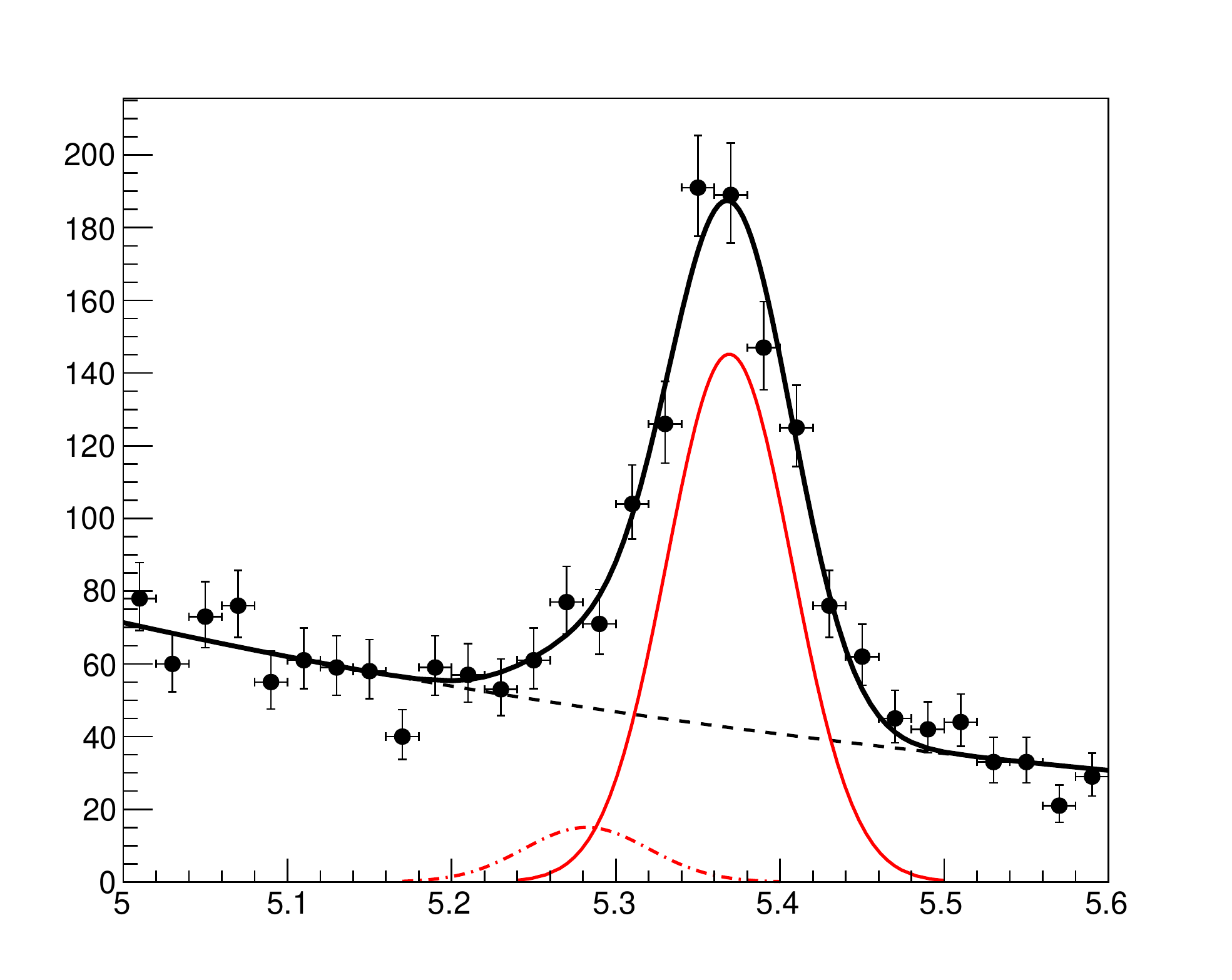}
    }
    \put(80,0){
      \includegraphics*[width=80mm,height=60mm%
      ]{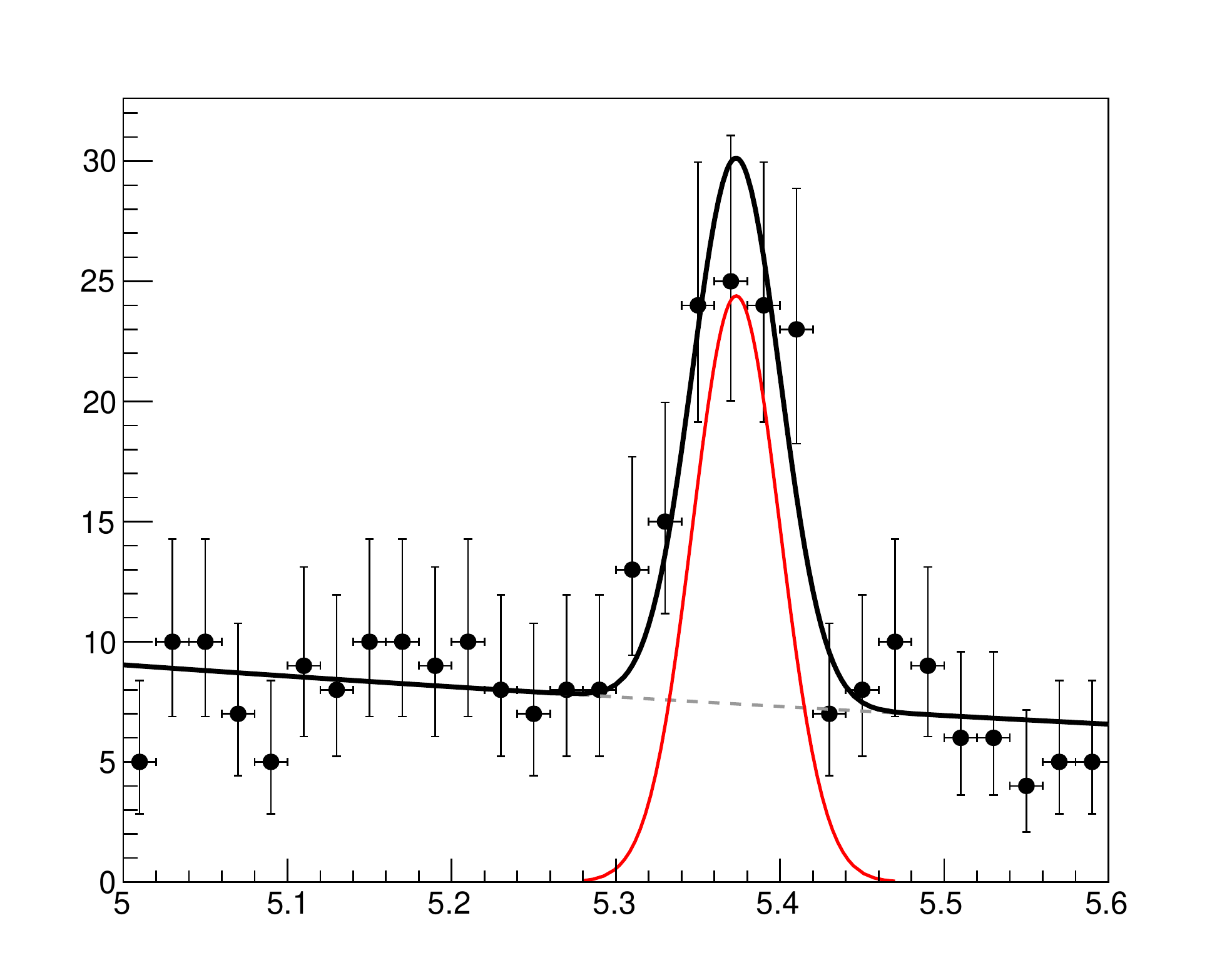}
    }
    \put(14,\lhcbypos){(a)}
    \put(94,\lhcbypos){(b)}
    \put( 0,13){\begin{sideways}\small{ Candidates/(20\mevcc) }\end{sideways}}    
    \put(80,13){\begin{sideways}\small{ Candidates/(20\mevcc) }\end{sideways}}    
    \put(\lhcbxpos,\lhcbypos){\small{\lhcb}}
    \put(\lhcbXpos,\lhcbypos){\small{\lhcb}}
    \put( 34,0){\small{M$(\jpsi\Peta)$}}
    \put(114,0){\small{M$(\psitwos\Peta)$}}
    \put( 59,0){\small{$\left[\gevcc\right]$}}
    \put(139,0){\small{$\left[\gevcc\right]$}}
  \end{picture}
  \caption {\small
    Mass distributions of
    (a)~$\BorBs\to\jpsi\Peta$ and (b)~$\BorBs\to\psitwos\Peta$ candidates. 
    The total fit function (solid black) and the combinatorial background (dashed) 
    are shown. The solid red lines show the signal \Bs~contribution and
    the red dot dashed line corresponds to the \Bd~contribution.
  }
  \label{fig:Fit_MB_ETA}
\end{figure}

The invariant mass distributions of the selected $\Ppsi\Peta$ candidates are shown
in Fig.~\ref{fig:Fit_MB_ETA}.
The $\Bs\to\Ppsi\Peta$ signal yields are estimated by performing 
unbinned extended maximum likelihood fits.
The $\Bs$ signal is modelled by a Gaussian distribution and the background
by an exponential function. 
In the $\jpsi\Peta$~case a possible contribution 
from the corresponding $\Bd$~decays is included 
in the fit model as an additional Gaussian component. 
The resolutions of the two Gaussian functions are set to be the same 
and the difference of their central 
values is fixed to the known difference between the $\Bs$ and the 
$\Bd$ masses~\cite{PDG2012}. The contribution from the decay $\Bd\to\psitwos\Peta$ is 
not considered in the baseline fit model. The mass resolution of the $\Bs\to\psitwos\Peta$ 
decay mode is fixed to the value 
$\sigma^{\psitwos\Peta}_{\rm DATA} =\sigma^{\jpsi\Peta}_{\rm DATA}\times\sigma^{\psitwos\Peta}_{\rm MC} / 
\sigma^{\jpsi\Peta}_{\rm MC}$, where $\sigma_{\rm DATA}$ and $\sigma_{\rm MC}$ are the 
widths of the corresponding channel in data and simulation, respectively.

The fit results are summarised in Table~\ref{tab:signal_fitres}.
In all cases the positions of the signal peaks are consistent with the nominal 
$\Bs$ mass~\cite{PDG2012} and the resolutions are in agreement with the
expectations from simulation. 
The measured yield of $\Bd\to\jpsi\Peta$ is $144\pm41$ events (uncertainty is 
statistical only), which is consistent with the expected value based on the measured 
branching fraction of this decay~\cite{Belle_BdJpsieta}. 
The statistical significance in each fit is determined as
\mbox{$\mathcal{S}=\sqrt{-2\ln{\frac{\mathcal{L}_\mathrm{B}}{\mathcal{L}_{\mathrm{S+B}}}}}$}
, where ${\mathcal{L}_{\mathrm{S+B}}}$ and ${\mathcal{L}_{\mathrm{B}}}$ denote the 
likelihood of the signal plus background hypothesis and the background only 
hypothesis, respectively. 
Taking into account the systematic uncertainty related to the fit function, which is 
discussed in detail in Sect.~\ref{sec:Efficiencies}, the significance of the 
$\Bs\to\psitwos\Peta$ signal is $6.2\sigma$. 

\begin{table}[t]
  \centering
  \caption{\small Fitted values of signal events~($N_{\mathrm{B}}$), signal peak 
  position~($\mathrm{M}_{\mathrm{B}}$) and resolution~($\sigma_{\mathrm{B}}$). 
  The quoted uncertainties are statistical only.
}\label{tab:signal_fitres}
\vspace*{3mm} 
\begin{tabular*}{0.75\textwidth}{@{\hspace{5mm}}l@{\extracolsep{\fill}}ccc@{\hspace{5mm}}}
  \multirow{2}*{~~~Mode}
  &  \multirow{2}*{$N_{\mathrm{B}}$}
  &  $\mathrm{M}_{\mathrm{B}}$ 
  &  $\sigma_{\mathrm{B}}$ 
  \\
  & 
  &  $\left[\mevcc\right]$ 
  &  $\left[\mevcc\right]$ 
  \\
  \hline
  $\Bs\to\jpsi\Peta$
  &  $   863\pm52 $   
  &  $5370.9\pm2.3$   
  &  $33.7  \pm2.3$ 
  \\
  $\Bs\to\psitwos\Peta$
  &  $\phantom{0}76\pm 12$  
  &  $5373.4\pm 5.0$  
  &   $26.6$~fixed
\end{tabular*}
\end{table}

\begin{figure}[ht]
  \setlength{\unitlength}{1mm}
  \centering
  \begin{picture}(160,60)
    \put(0,0){
      \includegraphics*[width=80mm,height=60mm%
      ]{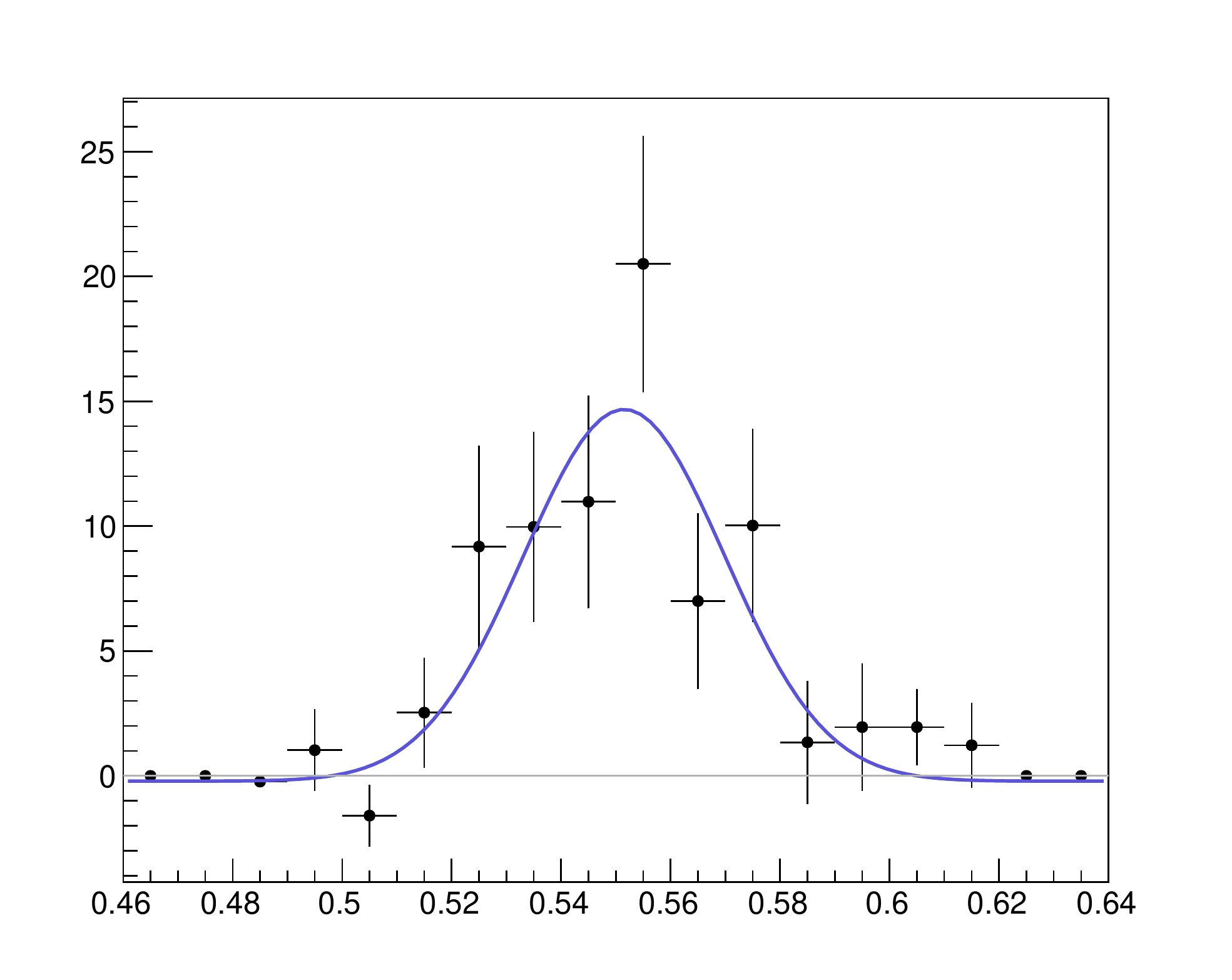}
    }
    \put(80,0){
      \includegraphics*[width=80mm,height=60mm%
      ]{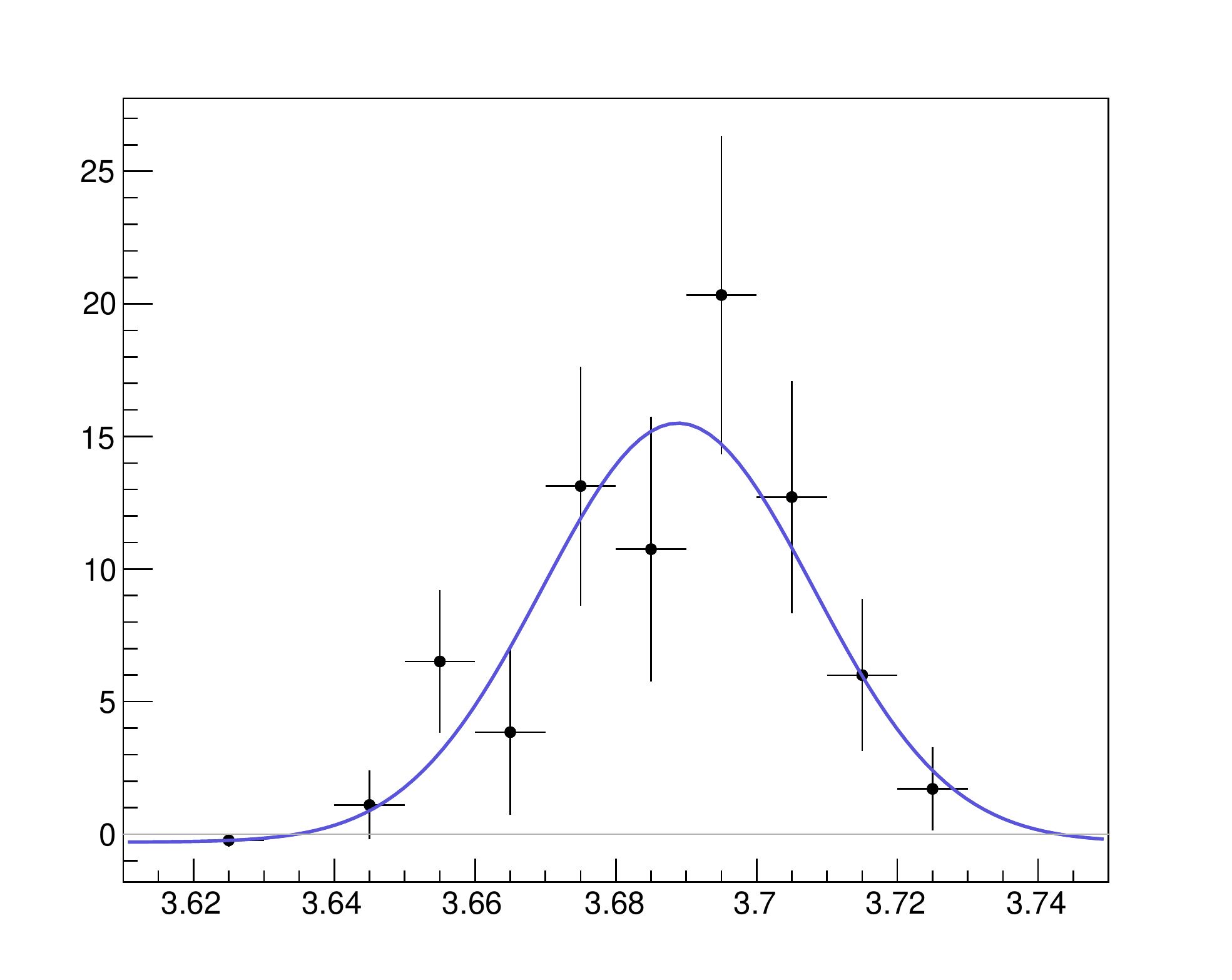}
    }
    \put(94,\lhcbypos){(b)}
    \put(14,\lhcbypos){(a)}
    \put( 34,0){\small{$\mathrm{M}(\Pgamma\Pgamma)$}}
    \put(114,0){\small{$\mathrm{M}(\mumu)$}}
    \put( 59,0){\small{$\left[\gevcc\right]$}}
    \put(139,0){\small{$\left[\gevcc\right]$}}
    \put(\lhcbxpos,\lhcbypos){\small{\lhcb}}
    \put(\lhcbXpos,\lhcbypos){\small{\lhcb}}
    \put( 0,13){\begin{sideways}\small{ Candidates/(10\mevcc)} \end{sideways}}    
    \put(80,13){\begin{sideways}\small{ Candidates/(10\mevcc)} \end{sideways}}    
  \end{picture}
  \caption {\small
    Background subtracted (a) $\Pgamma\Pgamma$ and (b) $\mumu$ 
    mass distributions in $\Bs\to\psitwos\Peta$ decays. 
    In both cases the blue line is the result of the fit described in the text.
  }
  \label{fig:Eta_reson}
\end{figure}
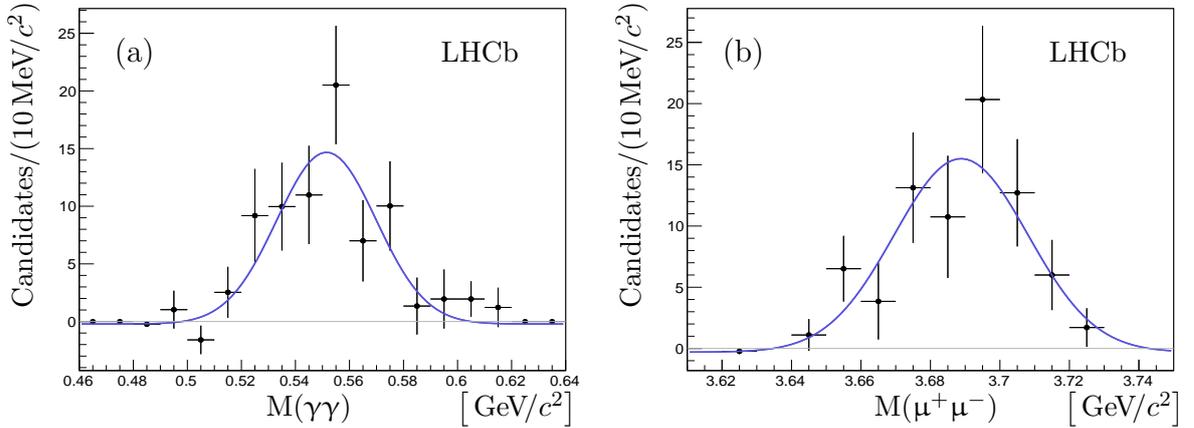

To demonstrate that the signal originates from $\Bs\to\psitwos\Peta$ decays 
the $\sPlot$~technique~\cite{Pivk:2004ty} has been used to separate the signal and the 
background. Using the $\mumu\Pgamma\Pgamma$ invariant mass distribution 
as the discriminating variable, the distributions for the invariant
masses of the intermediate resonances $\Peta\to\Pgamma\Pgamma$ 
and $\psitwos\to\mup\muM$ have been obtained. In this procedure, 
the invariant mass 
window for each corresponding resonance is released and the mass 
constraint is removed.
The resulting invariant mass distributions for $\Pgamma\Pgamma$ 
and $\mup\muM$ from $\Bs\to\psitwos\Peta$ candidates are shown 
in Fig.~\ref{fig:Eta_reson}. Clear signals are seen in both 
 $\Peta\to\Pgamma\Pgamma$ and $\psitwos\to\mup\muM$ decays. The 
distributions are described by the sum of a Gaussian function and a constant. 
The fit shows that the constant is consistent with zero, as expected.

\section[Observation of the          $\BorBs\to\psitwos\pipi$ decays]
        {Observation of the \boldmath$\BorBs\to\psitwos\pipi$ decays}
\label{subsec:B_to_psi_pipi}

\begin{figure}[b]
  \setlength{\unitlength}{1mm}
  \centering
  \begin{picture}(160,60)
    \put( 1,0){\includegraphics*[width=80mm,height=60mm%
      ]{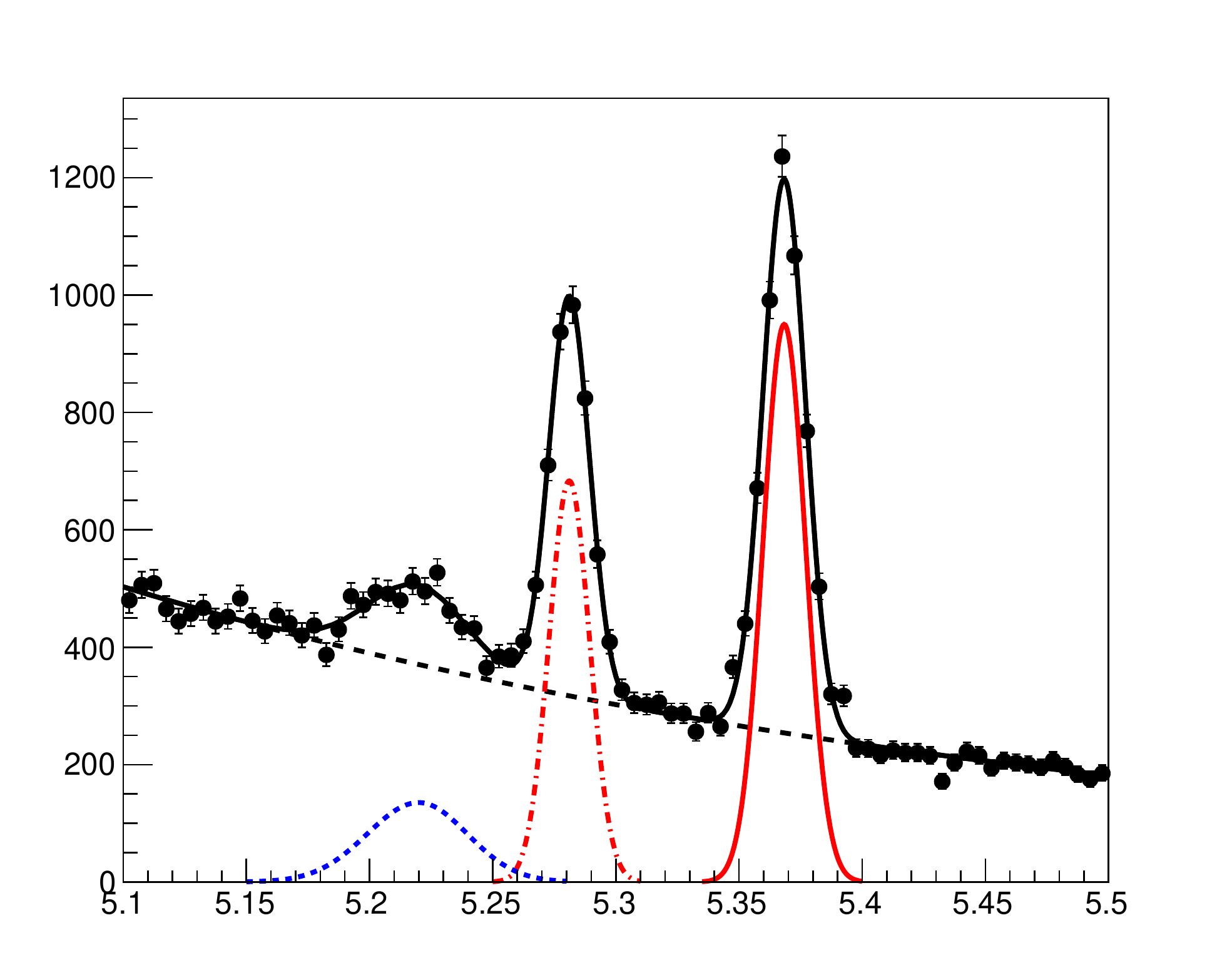}
    }
    \put(81,0){\includegraphics*[width=80mm,height=60mm%
      ]{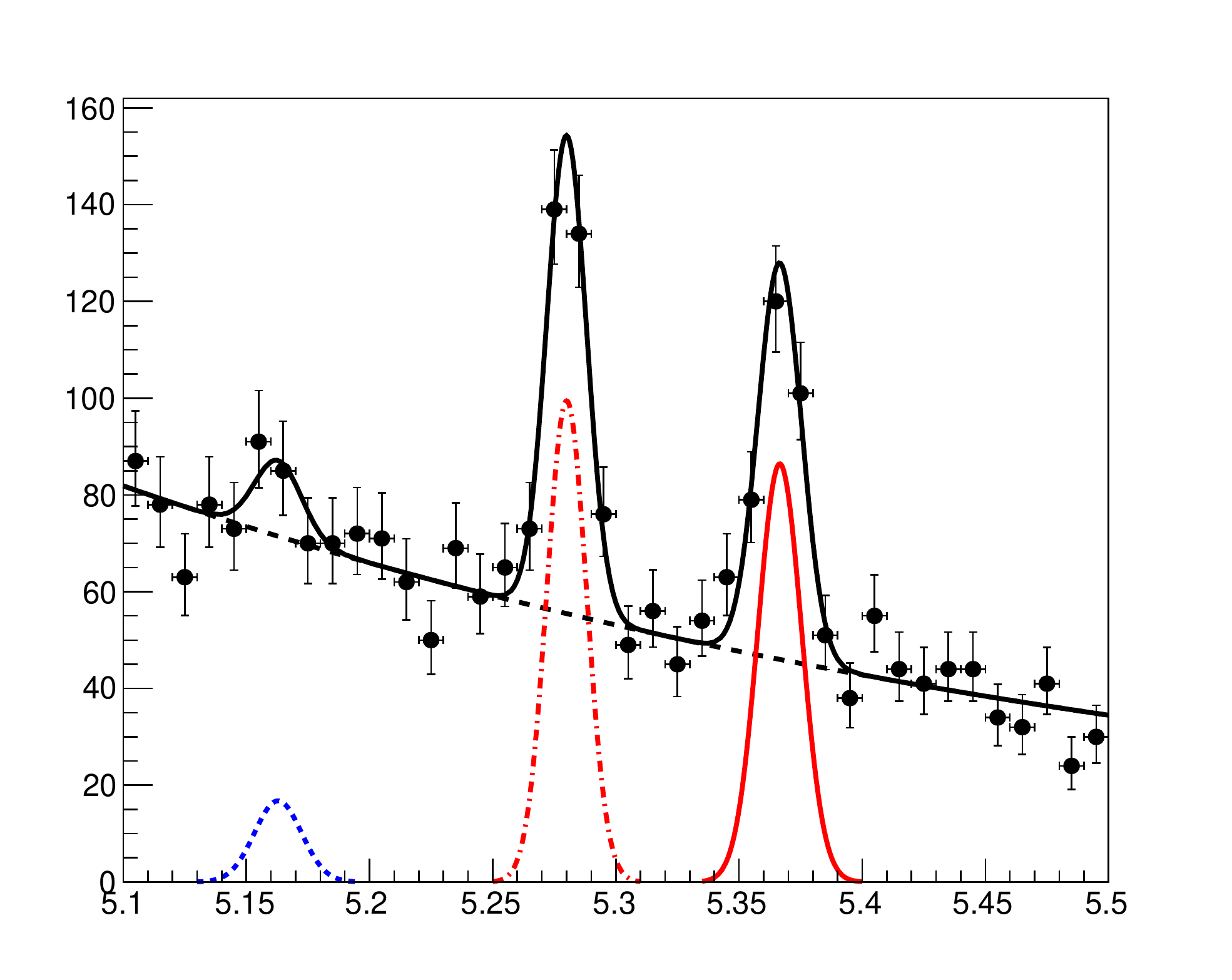}
    }
    \put(14,\lhcbypos){(a)}
    \put(94,\lhcbypos){(b)}
    \put(-1,13){\begin{sideways}\small{ Candidates/(5\mevcc) }\end{sideways}}    
    \put(80,13){\begin{sideways}\small{ Candidates/(10\mevcc) }\end{sideways}} 
    \put(\lhcbxpos,\lhcbypos){\small{\lhcb}}
    \put(\lhcbXpos,\lhcbypos){\small{\lhcb}}
    \put( 30,0){\small{$\mathrm{M}(\jpsi\pipi)$}}
    \put(110,0){\small{$\mathrm{M}(\psitwos\pipi)$}}
    \put( 59,0){\small{$\left[\gevcc\right]$}}
    \put(139,0){\small{$\left[\gevcc\right]$}}
  \end{picture}
  \caption {\small 
    Mass distributions of
    (a)~$\BorBs\to\jpsi\pipi$ and (b)~$\BorBs\to\psitwos\pipi$~candidates.
    The total fit 
    function (solid black) and the combinatorial background (dashed) are shown. 
    The solid red lines show the signal \Bs contribution and the red dot dashed 
    lines correspond to the \Bd contributions. The reflections from misidentified 
    $\Bd\to\Ppsi\Kstarz$, $\Kstarz\to\Kp\pim$ decays are shown with 
    dotted blue lines. 
  }
  \label{fig:Fit_MB_PIPI}
\end{figure}

The invariant mass distributions for the $\BorBs\to\Ppsi\pipi$ candidates are 
shown in Fig.~\ref{fig:Fit_MB_PIPI}. The narrow signals correspond to 
the \mbox{$\Bd\to\Ppsi\pipi$} and $\Bs\to\Ppsi\pipi$ decays. The peak at lower 
mass corresponds to 
a reflection 
from $\Bd\to\Ppsi\Kstarz(\to\Kp\pim)$ decays 
where the kaon is misidentified as a pion. The contribution from 
$\Bs\to\Ppsi\Kstarz$~decays~\cite{LHCb-PAPER-2012-014} is negligible.

The invariant mass distributions are fitted with two Gaussian 
functions to describe the two signals, an asymmetric Gaussian function 
with different width for the two sides to represent the reflection from 
$\Bd\to\Ppsi\Kstarz$ decays and an exponential function for the background. 
The fit results are summarised in Table~\ref{tab:signal_fitres_pipi}.
The statistical significances of the signals are found to be 
larger than 9 standard deviations.

\begin{table}[tb]
  \centering
\caption{\small Fitted values of signal events~($N_{\mathrm{B}}$), signal peak 
  position~($\mathrm{M}_{\mathrm{B}}$) and resolution~($\sigma_{\mathrm{B}}$). 
  The quoted uncertainties are statistical only.
}\label{tab:signal_fitres_pipi}
\vspace*{3mm} 
\begin{tabular*}{0.75\textwidth}{@{\hspace{5mm}}l@{\extracolsep{\fill}}ccc@{\hspace{5mm}}}
  \multirow{2}*{~~~Mode}
  &  \multirow{2}*{$N_{\mathrm{B}}$}
  &  $\mathrm{M}_{\mathrm{B}}$ 
  &  $\sigma_{\mathrm{B}}$ 
  \\
  & 
  &  $\left[\mevcc\right]$ 
  &  $\left[\mevcc\right]$ 
  \\
  \hline
  $\Bd\to\jpsi\pipi$ 	
  &  $2801   \pm 85$  
  &  $5281.1 \pm 0.3$  
  &  $8.2    \pm 0.3$
  \\
  $\Bs\to\jpsi\pipi$ 	
  &  $4096   \pm 86$  
  &  $5368.4 \pm 0.2$  
  &  $8.7    \pm 0.2$ 
  \\
  $\Bd\to\psitwos\pipi$ 	
  &  $\phantom{0}202 \pm 23$ 	
  &  $5280.3 \pm 1.0$  
  &  $8.4    \pm 1.1$
  \\
  $\Bs\to\psitwos\pipi$ 	
  &  $\phantom{0}178 \pm 22$ 	
  &  $5366.3         \pm 1.2$  
  &  $9.1            \pm 1.4$
\end{tabular*}
\end{table}

%

\begin{figure}[t]
  \setlength{\unitlength}{1mm}
  \centering
  \begin{picture}(160,60)
    \put( 1,0){
      \includegraphics*[width=80mm,height=60mm%
      ]{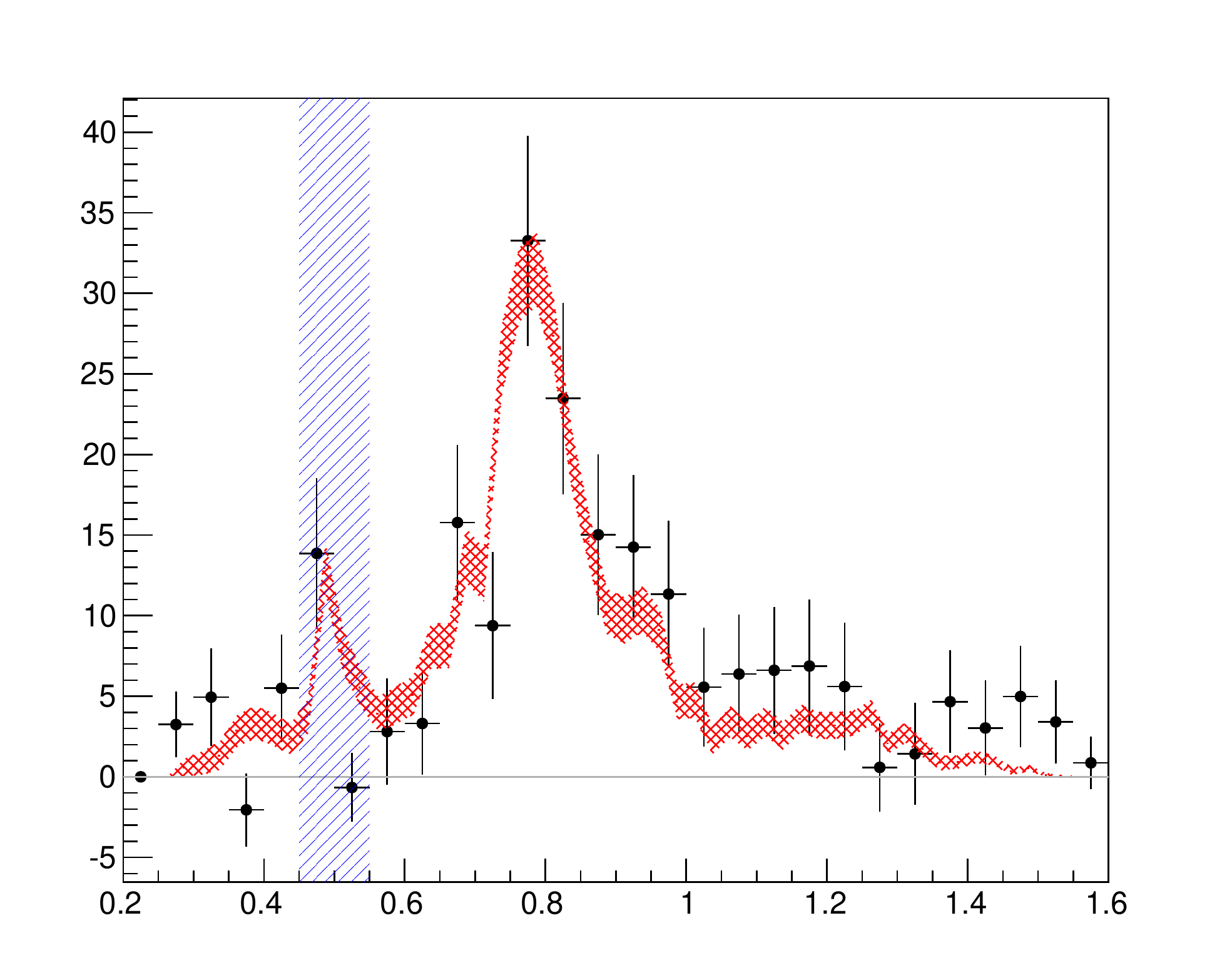}
    }
    \put(81,0){\includegraphics*[width=80mm,height=60mm%
      ]{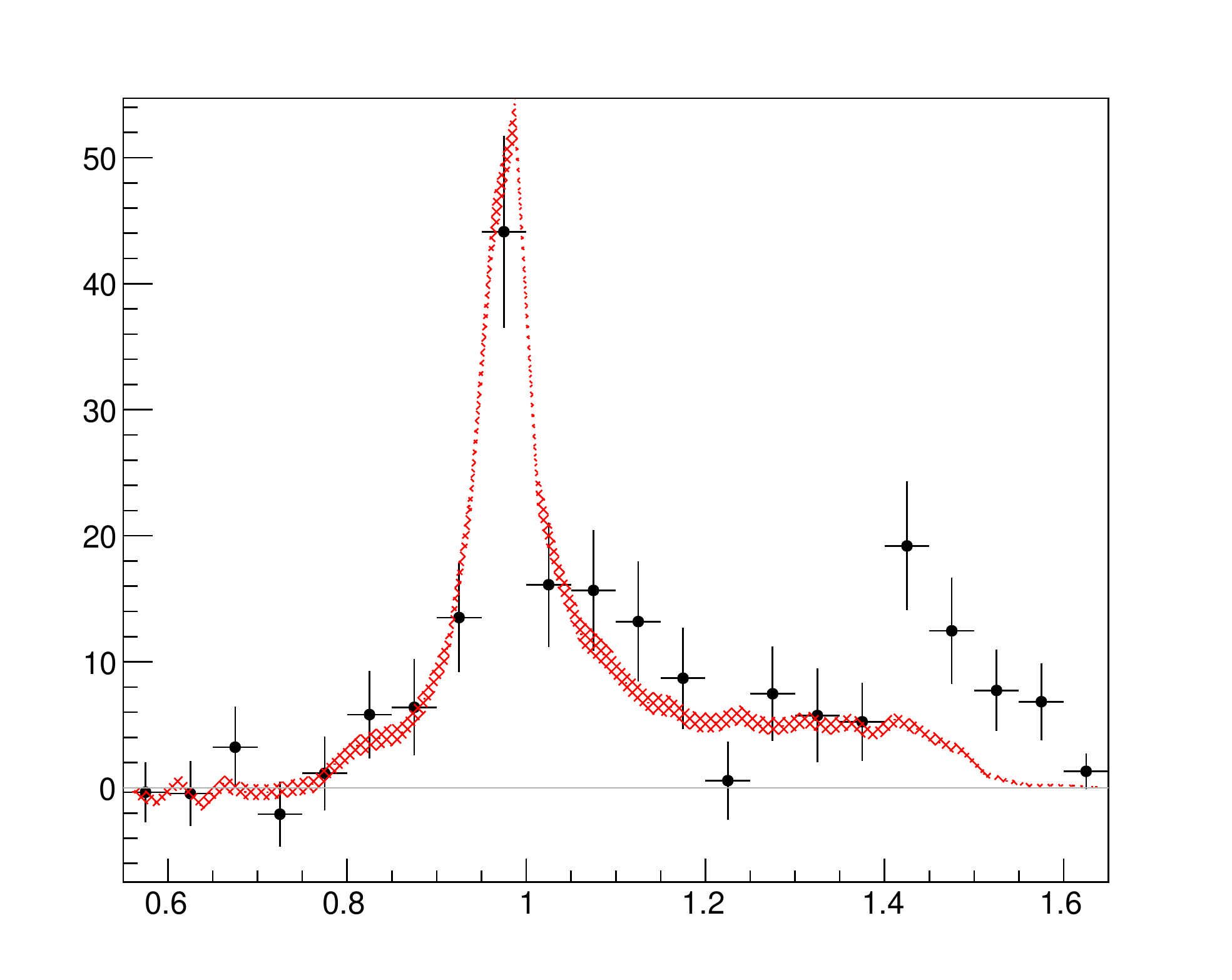}
    }
    \put(14,\lhcbypos){(a)}
    \put(94,\lhcbypos){(b)}
    \put( 0,13){\begin{sideways}\small{ Candidates/(50\mevcc) }\end{sideways}}    
    \put(80,13){\begin{sideways}\small{ Candidates/(50\mevcc) }\end{sideways}} 
    \put(\lhcbxpos,\lhcbypos){\small{\lhcb}}
    \put(\lhcbXpos,\lhcbypos){\small{\lhcb}}
    \put( 33,0){\small{$\mathrm{M}(\pipi)$}}
    \put(113,0){\small{$\mathrm{M}(\pipi)$}}
    \put( 59,0){\small{$\left[\gevcc\right]$}}
    \put(139,0){\small{$\left[\gevcc\right]$}}
  \end{picture}
  \caption {\small
    Background subtracted \pipi~mass distribution in
    (a)~$\Bd\to\psitwos\pipi$ and (b)~$\Bs\to\psitwos\pipi$ (black points). 
    The red filled area shows 
    the expected signal spectrum for the \psitwos channel derived from the measured 
    spectrum of the \jpsi channel 
    (the fit has one parameter --- the normalisation). The width of the band corresponds 
    to the uncertainties of the distribution from the \jpsi channel. In case of 
    $\Bd\to\psitwos\pipi$, the blue vertical filled area shows the \KS region that 
    is excluded from the fit.
  }
  \label{fig:PIPI_DISTS}
\end{figure}

For the~$\BorBs\to\jpsi\pipi$ decays, the \pipi mass shapes have been 
studied in detail using a partial wave analysis 
in Refs.~\cite{LHCb-PAPER-2012-045,LHCb-PAPER-2012-005}. The main contributions are 
$\Bd\to\jpsi\rhomeson$ and $\Bs\to\jpsi \mathrm{f_0(980)}$. However, due to the limited 
number of signal events, 
the same method cannot be used for the $\BorBs\to\psitwos\pipi$ decays.
The $\sPlot$~technique is used in order to study the dipion mass 
distribution in those decays. 
With the $\psitwos\pipi$ invariant mass as the discriminating variable, the \pipi 
invariant mass spectra from $\BorBs\to\psitwos\pipi$ decays are obtained (see Fig.~\ref{fig:PIPI_DISTS}).

To check that the background subtracted \pipi~distributions have similar shapes in 
both channels, the distribution obtained from the $\psitwos\pipi$ decay is fitted with 
the distribution obtained from the $\jpsi\pipi$ channel, 
corrected by the ratio of phase-space factors and by the ratio of the efficiencies 
which depends on the dipion invariant mass. 
The p-value for the $\chi^2$~fit is 30\% for $\Bd\to\Ppsi\pipi$ 
and 7\% for $\Bs\to\Ppsi\pipi$, respectively. 
As seen in Fig.~\ref{fig:PIPI_DISTS}, 
$\Bd\to\psitwos\rhomeson$ and 
$\Bs\to\psitwos\mathrm{f_0(980)}$~decays are 
the main contributions to $\BorBs\to\psitwos\pipi$ decays.
Detailed amplitude analyses of the resonance structures in 
$\BorBs\to\psitwos\pipi$ decays, similar to 
Refs.~\cite{LHCb-PAPER-2012-045,LHCb-PAPER-2012-005}, 
will be possible with a larger dataset. 
This  will allow the possible excess of events 
in the region $\mathrm{M}(\pipi)>1.4\gevcc$ to be investigated.

The narrow peak around $0.5\gevcc$ in Fig.~\ref{fig:PIPI_DISTS}(a) is dominated
by $\KS\to\pipi$ from $\Bd\to\jpsi\KS$ decays. The contributions 
from \KS~decays are taken into account by the fit function described in 
Ref.~\cite{LHCb-PAPER-2012-022}. The resulting yields are 
$129\pm26$ in the $\jpsi$ channel and $11\pm6$ in the $\psitwos$ channel. 
In the calculation of the final ratio of branching fractions, the number 
of \KS~events is subtracted from the corresponding $\Bd\to\Ppsi\pipi$~yields. 
The yield from $\Bs\to\Ppsi\KS$~decays is negligible~\cite{LHCb-PAPER-2011-041}.

%% file: efficiency.tex
\section{Efficiencies and systematic uncertainties}
\label{sec:Efficiencies}

The ratios of branching fractions are calculated using the formula 
\begin{equation}
\frac{\BR({\mathrm{\B}\to\psitwos\mathrm{X^{0}}})}{\BR({\mathrm{B}
\to\jpsi\mathrm{X^{0}}})} = \frac{{N}_{\psitwos \mathrm{X^{0}}}}{{N_{\jpsi
X^{0}}}} \times \frac{\mathrm{\epsilon_{\jpsi X^{0}}}}{\mathrm{\epsilon}_{\psitwos \mathrm{X^{0}}}}
\times \frac{\mathrm{\BR(\jpsi \to \mup \muM)}}{\BR(\psitwos \to \mup \muM)}\;\mathrm{,}
\label{eq:overall}
\end{equation}
where ${N}$ is the number of signal events, and $\mathrm{\epsilon}$ is the 
product of the geometrical acceptance, the detection, reconstruction, 
selection and trigger efficiencies. 
The efficiency ratios are estimated using simulation for all six decay modes.

The efficiency ratios are $1.22\pm0.01$, $1.03\pm0.01$ and $1.02\pm0.01$ for the 
$\Bs\to\Ppsi\Peta$, $\Bd\to\Ppsi\pipi$ and $\Bs\to\Ppsi\pipi$ 
channels, respectively (uncertainties are statistical only). Since the selection 
criteria for the decays with \jpsi and 
\psitwos are identical, the ratio of efficiencies is expected to be close to 
unity. The deviation of the overall efficiency ratio from unity in case of
$\Bs\to\Ppsi\Peta$ is due to the difference
between the \pt spectra of the selected \jpsi and \psitwos mesons,
when the $\pt(\Peta)>2.5\gevc$~requirement is applied.
For the $\BorBs\to\Ppsi\pipi$ channels this 
effect is small since no explicit \pt requirement is applied
on the dipion system.

Most systematic uncertainties cancel in the ratio of branching fractions, in particular, 
those related to the muon and \Ppsi reconstruction and identification.
Systematic uncertainties related to the fit model are estimated using
 a number of alternative models for the description of the invariant mass
 distributions. For the $\Bs\to\Ppsi\Peta$ decays the tested alternatives 
 are a fit model including a \Bd signal component (with the ratio 
 $N(\Bd\to\Ppsi\Peta$)/$N(\Bs\to\Ppsi\Peta$) fixed from the \jpsi channel), 
a fit model with a linear function for the background
 description, fits with signal widths fixed or not fixed to those obtained in 
 simulation, a 
fit with the difference between the fitted \Bd and \Bs~masses allowed to vary 
 within a $\pm1\sigma$ interval around the nominal value~\cite{PDG2012}, and a fit model
 with Student's t--distributions for the signals. 
For each alternative fit
model the ratio of event yields is calculated and the systematic uncertainty 
is then determined as the maximum deviation of this ratio from the ratio 
obtained with the baseline model. For $\BorBs\to\Ppsi\pipi$ decays the tested 
alternatives include a fit with a first or second order polynomial for the background
description, a model with a symmetric Gaussian distribution for the reflection and
 a model with the difference of the mean values of the two Gaussian functions 
 fixed to the known mass difference between the $\Bs$ and the $\Bd$ mesons~\cite{PDG2012}.
The maximum deviation observed in the ratio of yields in the $\psitwos$ and $\jpsi$ modes
is taken as the systematic uncertainty. The obtained uncertainties are $8.0\%$ for
the $\Bs\to\Ppsi\Peta$ channel, $1.0\%$ for the $\Bd\to\Ppsi\pipi$ channel and $1.6\%$
for the $\Bs\to\Ppsi\pipi$ channel.

The selection efficiency for the dipion system has a dependence
on the dipion invariant mass. 
The ratios of efficiencies vary 
over the entire \pipi mass range
by approximately 40\% and 24\% for $\Bd\to\Ppsi\pipi$ 
and $\Bs\to\Ppsi\pipi$~channels, respectively. 
The systematic uncertainties related 
to the different dependence of the efficiency as a function of the dipion invariant 
mass for \jpsi and \psitwos channels are evaluated using the decay models from 
Ref.~\cite{LHCb-PAPER-2012-005} for \Bs and 
Refs.~\cite{LHCb-PAPER-2012-022,LHCb-PAPER-2012-045} for \Bd decays. 
The systematic uncertainties on the branching fraction ratios are 2\% for both channels. 

The most important source of uncertainty arises from potential disagreement between
data and simulation in the estimation of efficiencies. 
This source of uncertainty is studied by varying the selection 
criteria in ranges corresponding to approximately $15\%$ change in the signal yields. 
The agreement is estimated by comparing the efficiency corrected ratio of yields 
with these variations.
The resulting uncertainties are found to be 11.5\% in the $\Bs\to\Ppsi\Peta$ channel
and 8\% in the \mbox{$\BorBs\to\Ppsi\pipi$} channel. 

The geometrical acceptance is calculated separately for different magnet 
polarities. The observed difference in the efficiency ratios 
is taken as an estimate of the systematic uncertainty and is $1.1\%$ for 
the $\Bd\to\Ppsi\pipi$ channel and negligible for the other channels.

The trigger is highly efficient in selecting $\mathrm{B}$ meson decays with two muons in the final
state. For this analysis the dimuon pair is required to trigger the event. Differences in
the trigger efficiency between data and simulation are studied in the data using events
that were triggered independently of the dimuon pair~\cite{LHCb-PAPER-2012-010}. Based on these studies, an
uncertainty of 1.1\% is assigned.
A summary of all systematic uncertainties is presented in Table~\ref{table:SYSTEMATICS}.

\begin{table}[ht]
  \centering 
  \caption{\small Relative systematic uncertainties (in \%) of the relative branching fractions.}
  \label{table:SYSTEMATICS}
  \vspace*{3mm} 
  \begin{tabular*}{0.90\textwidth}{@{\hspace{5mm}}l@{\extracolsep{\fill}}ccc@{\hspace{5mm}}}
      Source  
      & $\Bs\to\Ppsi\Peta$
      & $\Bd\to\Ppsi\pipi$ 
      & $\Bs\to\Ppsi\pipi$ 
      \\
      \hline
      Fit model 
      &  \phantom{0}8.0     
      &  1.0   
      &  \phantom{0}1.6
      \\
      Mass dependence of efficiencies
      &  --- 
      &  2.0  
      &  \phantom{0}2.0 
      \\
      Efficiencies from simulation 
      &  11.5 
      &   8.0 
      &  \phantom{0}8.0
      \\
      Acceptance 
      &  $\hspace{-2.5mm}<0.5$ 
      &  1.1 
      &  $\hspace{-2.5mm}<0.5$ 
      \\ 
      Trigger 
      &  \phantom{0}1.1	
      &  1.1	
      &  \phantom{0}1.1	
      \\
      \hline
      Sum in quadrature 
      &  14.1 
      &   8.5 
      &  \phantom{0}8.5
  \end{tabular*}
\end{table}

%% file: results.tex
\section{Results}
\label{sec:Result}

With data corresponding to an integrated luminosity of 1.0\invfb, 
collected in 2011 with the \lhcb detector, 
the first  observations of the $\Bs\to\psitwos\Peta$~and 
$\BorBs\to\psitwos\pipi$~decays have been made. 
The relative rates of \BorBs~meson decays into final states 
containing \jpsi~and \psitwos~mesons are measured for 
those decay modes. 
Since the dielectron branching fractions of \Ppsi~mesons are 
measured more precisely than those of the dimuon decay modes, 
invoking lepton universality, the ratio 
$\frac{\BR(\jpsi\to\mup\muM)}{\BR(\psitwos\to\mup\muM)}=\frac{\BR(\jpsi\to\ep\en)}{\BR(\psitwos\to\ep\en)}=7.69\pm0.19$~\cite{PDG2012} 
is used. 
The results are combined using Eq.~\eqref{eq:overall}, to give
\begin{align*}
\frac{\BR(\Bs\to \psitwos \Peta)}{\BR(\Bs\to \jpsi \Peta)} &=0.83\pm0.14\,\stat\pm0.12\,\syst\pm0.02\,(\BR), \\
\frac{\BR(\Bd\to \psitwos \pipi)}{\BR(\Bd\to \jpsi \pipi )}&=0.56\pm0.07\,\stat\pm0.05\,\syst\pm0.01\,(\BR), \\
\frac{\BR(\Bs\to \psitwos \pipi)}{\BR(\Bs\to \jpsi \pipi)} &=0.34\pm0.04\,\stat\pm0.03\,\syst\pm0.01\,(\BR), 
\end{align*}
where the first uncertainty is statistical, the second systematic and the third from 
the world average ratio~\cite{PDG2012} of the \jpsi~and \psitwos~branching fractions to 
dileptonic final states. The branching fraction ratios measured here correspond 
to the time integrated quantities. 
For the $\Bd\to\jporptwo\pipi$~channel 
the measured ratio excludes the $\KS\to\pipi$~contibution. 
The dominant contributions to the 
$\BorBs\to\psitwos\pipi$ decays are found to be from $\Bd\to\psitwos\rhomeson$ and 
$\Bs\to\psitwos\mathrm{f_0(980)}$ decays.

These results  
are compatible with the measured range of relative branching 
fractions of B decays to \psitwos and \jpsi mesons. The $\Bs\to\psitwos\Peta$ 
and $\Bs\to\psitwos\pipi$ decays are particularly interesting since, with more 
data becoming available, they can be used to measure $\CP$ violation in \Bs mixing.
\section*{Acknowledgements}
\noindent We express our gratitude to our colleagues in the CERN
accelerator departments for the excellent performance of the LHC. We
thank the technical and administrative staff at the LHCb
institutes. We acknowledge support from CERN and from the national
agencies: CAPES, CNPq, FAPERJ and FINEP (Brazil); NSFC (China);
CNRS/IN2P3 and Region Auvergne (France); BMBF, DFG, HGF and MPG
(Germany); SFI (Ireland); INFN (Italy); FOM and NWO (The Netherlands);
SCSR (Poland); ANCS/IFA (Romania); MinES, Rosatom, RFBR and NRC
``Kurchatov Institute'' (Russia); MinECo, XuntaGal and GENCAT (Spain);
SNSF and SER (Switzerland); NAS Ukraine (Ukraine); STFC (United
Kingdom); NSF (USA). We also acknowledge the support received from the
ERC under FP7. The Tier1 computing centers are supported by IN2P3
(France), KIT and BMBF (Germany), INFN (Italy), NWO and SURF (The
Netherlands), PIC (Spain), GridPP (United Kingdom). We are thankful
for the computing resources put at our disposal by Yandex LLC
(Russia), as well as to the communities behind the multiple open
source software packages that we depend on.
\clearpage